\documentstyle[psfig]{mn}
\newif\ifAMStwofonts
\AMStwofontstrue

\def\h2{H$_2$}

\def\hii{H{\sevensize II}}

\def\kms{$\mbox{km~s}^{-1}$}

\def\mic{$\,\mu$m}
\def\cmt{{$\rm \,cm^{-3}$}}
\def\cmtwo{{$\rm \,cm^{-2}$}}

\title[Formation Pumping of H$_2$ in M\,17]
{Formation Pumping of Molecular Hydrogen in the Messier 17
Photodissociation Region}
\author[M.G. Burton et al.]
       {M.\ G. Burton$^{1,2}$\thanks{email: M.Burton@unsw.edu.au}, 
	D.\ Londish$^{1,3}$ and
        P.\ W.\ J.\ L.\ Brand$^{4}$ \\
$^1$ School of Physics, University of New South Wales, Sydney, 
NSW 2052, Australia\\
$^2$ School of Cosmic Physics, Dublin Institute for Advanced Studies, 
5 Merrion Square, Dublin 2, Ireland \\
$^3$ School of Physics, University of Sydney, Sydney NSW 2006, Australia \\
$^4$ Institute for Astronomy, University of Edinburgh, Royal Observatory, 
Blackford Hill, Edinburgh, EH9 3HJ}

\date{Accepted, Received, In original form 4 December 2001.}

\pagerange{\pageref{firstpage}--\pageref{lastpage}}
\pubyear{2002}

\begin{document}

\maketitle

\label{firstpage}

\begin{abstract}  

We have imaged the emission from the near-infrared v=1--0 S(1), 1--0
S(7), 2--1 S(1) and 6--4 O(3) lines of molecular hydrogen in the N--
and SW--Bars of M\,17, together with the hydrogen Br$\gamma$ and Br10
lines.  This includes the first emission line image ever to be
obtained of a line from the highly excited v=6 level of H$_2$.  In
both Bars, the H$_2$ emission is generally distributed in clumps along
filamentary features.  The 1--0 S(1) and 2--1 S(1) images have similar
morphologies. Together with their relative line ratios, this supports
a fluorescent origin for their emission, within a photodissociation
region.  The SW--Bar contains a clumpy medium, but in the N--Bar the
density is roughly constant. The 1--0 S(7) line image is also similar
to the 1--0 S(1) image, but the 6--4 O(3) image is significantly
different to it.  Since the emission wavelengths of these two lines
are similar (1.748 to 1.733\mic), this cannot be due to differential
extinction between the v=6 and the v=1 lines.  We attribute the
difference to the pumping of newly formed H$_2$ into the v=6, or to a
nearby, level.  However, this also requires either a time-dependent
photodissociation region (where molecule formation does not balance
dissociation), rather than it to be in steady-state, and/or for the
formation spectrum to vary with position in the source.  If this
interpretation of formation pumping of molecular hydrogen is correct,
it is the first clear signature from this process to be seen.

\end{abstract}

\begin{keywords}
stars: formation -- ISM: molecules -- infrared: ISM -- molecular processes --
ISM: \hii\ regions -- ISM: individual: M\,17
\end{keywords}

\section{INTRODUCTION}
\subsection{Messier 17}
Messier 17 is one of the nearest regions of massive star formation to
us, situated $\sim 1.3$\,kpc away (Hanson, Howarth \& Conti 1997).  It
is particularly notable for the bright \hii\ region which attracted
Messier's attention (and also known as the Omega Nebula). Within it
lies an obscured cluster of young stars. Thirteen OB stars have been
spectroscopically identified (Hanson et al.\ 1997), all with between 4
and 15 magnitudes of extinction at optical wavelengths and estimated
to be $\sim 10^6$\,years old.  The \hii\ region is adjacent to a
molecular cloud complex, in particular the core M\,17 SW, which
contains more than $\rm 10^4 M_{\sun}$ of material (Lada 1976,
Thronson \& Lada 1983).  Intense far--IR emission arises from the
core, with a luminosity of $\rm \sim 6 \times 10^6 L_{\sun}$ (Harper
et al.\ 1976, Wilson et al.\ 1979). While this originates from dust
exposed to far--UV radiation, the core appears to be externally
heated, by the stars that excite the
\hii\ region (Gatley et al.\ 1979), rather than internally.  It forms
a photodissociation region (PDR, Tielens \& Hollenbach 1985).  The
method of excitation of the hydrogen molecules in it are the subject
of this paper.

\subsection{Molecular Hydrogen Observations in M\,17}
M\,17 was first mapped in H$_2$ through the 2.12\mic\ 1--0 S(1) line
by Meadows (1986) and Gatley \& Kaifu (1987). Their maps reveal two
obvious regions of emission, a bar running across the north, and
another running south to west.  They are known as the `Northern Bar'
and the `South Western Bar', respectively.  They closely follow two
emission ridges seen in radio continuum from 1.3 to 21\,cm (Felli et
al.\ 1984).  These bars are PDRs, interface regions between the
ionized and the molecular gas.  The Northern Bar also follows an
optically visible bar of the \hii\ region, suggesting that the
extinction to it is relatively small.  The South Western Bar, however,
is optically obscured. PDR modelling of far--IR and sub--mm line
emission features from the SW--Bar (Burton, Hollenbach \& Tielens
1990, Meixner et al.\ 1992) indicates that it is a clumpy region,
containing gas with a mix of densities, ranging from $10^3$ to
$10^7$\,\cmt, exposed to a far--UV radiation field of $\sim 10^4$
times the average interstellar radiation field.

Low spatial resolution (20 arcsec beam) near--IR spectroscopy of the
1--0 and 2--1 S(1) H$_2$ lines (Tanaka et al.\ 1989) in the N--Bar
found a line ratio of $\sim 2$.  This is the ratio expected if the
excitation is via a fluorescent cascade through the
vibrational-rotational levels of the ground electronic state (Black \&
Dalgarno 1976). The molecule is first excited by far--UV photons to an
excited electronic level, from which it decays.  Chrysostomou et al.\
(1992, 1993) imaged the N--Bar with arcsecond spatial resolution in
the H$_2$ 1--0 S(1) line, as well as in the 3.29\mic\ PAH emission
feature and the hydrogen Br$\gamma$ line at 2.17\mic.  They also
obtained long slit spectroscopy across the N--Bar from 2.0--2.5\mic,
measuring 16 lines of H$_2$, as well as several
\hii\ region emission lines.  The spatial coincidence found for the PAH
and H$_2$ emission features confirmed the fluorescent origin, and also
suggested that most of the gas had density $> 10^5$\cmt.  They found
the 1--0/2--1 S(1) ratio to be $\sim 3$, and to be constant along the
60\,arcsec long slit.  This not only suggests that the H$_2$ is
slightly thermalised (i.e.\ the density is about the critical density,
$\sim 10^5$\cmt), but that the physical conditions along the N--Bar
are constant.

\subsection{Molecular Hydrogen Excitation in PDRs}
In dense PDRs, with a strong far--UV radiation field, self-shielding
can bring the H/H$_2$ dissociation front to optical depths $A_v <
1$\,mag from the ionized surface of the cloud.  Here the temperature
can reach 1,000--2,000\,K, allowing the v=1 level of H$_2$ to be
thermalised.  H$_2$ line ratios, for instance the 1--0/2--1 S(1)
ratio, then increase from the pure fluorescent value of 2, towards the
thermal value of 10 observed in shocks and in gas at $\sim 2000$\,K
(see Sternberg \& Dalgarno 1989, Burton, Hollenbach \& Tielens 1990).
The actual line ratio depends on the density of the gas. If this is
not constant, then the ratio will likely vary with position in a
source.  However calculating its value depends on an accurate
knowledge of the collisional excitation rate coefficients for H$_2$,
which are poorly known.  Hence the quantitative application of
``collisional fluorescence'' models to the interpretation of H$_2$
lines ratios in PDRs requires some care.

A component of the fluorescent emission from H$_2$ must arise due to
the formation of the hydrogen molecules themselves.  For
photo-excitation of H$_2$ molecules by far--UV photons, approximately
15\% lead to dissociation.  However the hydrogen molecules can be
reformed, through the recombination of two hydrogen atoms on the
surfaces of dust grains (Hollenbach \& Salpeter 1971).  In the steady
state, the rate of formation of H$_2$ molecules balances its rate of
photodissociation.  The 4.5\,eV bond energy of H$_2$ is released on
formation.  Some of this is used in escaping from the surface of the
dust grain, some goes into translational kinetic energy in the
molecule, and the rest goes into internal energy.  The new molecules
can then emerge in a excited vibrational-rotational state, from which
radiative decay will occur.  Thus a formation component to the H$_2$
spectrum is expected.  However, the signature of this spectrum is
unclear.

As Black \& Dalgarno (1976) first realised, when calculating the
infrared spectrum of fluorescent H$_2$, the contribution from molecule
formation may be important.  In their original model they assumed
equipartition of the binding energy released on formation.  They
divided this equally between the internal energy of the molecule, its
translational energy on escape from the grain surface, and the
internal energy imparted to the grain lattice.  They further assumed
that the 1.5\,eV provided as internal energy to the molecule was
spread with a Boltzmann distribution through the
vibrational-rotational levels.  However, other assumptions are
possible.  For instance, Hunter \& Watson (1978) argue that H$_2$
molecules are released in rotationally hot, vibrationally cold states
(i.e.\ high--J $\geq 7$, low--v). A model by Duley \& Williams (1986),
on the other hand, although agreeing that the molecules should form
hot, argued that they would appear in just the opposite combination of
states (i.e.\ low--J, high--v, with v=6 likely).  Le Bourlot et al.\
(1995) have investigated the infrared spectrum produced for pure
formation pumping of H$_2$, under a variety of formation models.  For
instance, as might be expected, the intensities of v=6 lines are found
to be considerably greater under the Duley \& Williams (1986) model
than in other models.  However, since this model does not include the
fluorescent cascade component to the emission, which will generally
dominate the intensity of the H$_2$ lines, it is hard to use it to
undertake a quantitative comparison with data.

\subsection{Previous Reports of the Detection of Formation Pumping for H$_2$}
There have been few reports made of the signature for H$_2$ formation
being observed in a spectrum. Wagenblast (1992) considers UV
absorption spectra of three nearby diffuse clouds, along lines of
sight to background stars. He finds that the populations in the
excited, pure-rotational levels, v=0, J=5, 6 \& 7, along the sight
lines, cannot be produced by UV or thermal excitation.  Assuming that
molecule formation takes place in two adjacent rotational levels
within a particular vibrational state (one for ortho-H$_2$, or odd--J,
and the other for para-H$_2$, or even--J), he calculates the possible
pairings that could account for the observed ratios of the three
lines. He found that H$_2$ would need to be formed in a rotationally
hot state ($\rm J \geq 7$), but with a range of vibrational states
possible, up to v=11. Federman et al.\ (1995) explore this further
with additional absorption measurements from the v=3 level in the
source $\zeta$\,Oph, though do not find any fit to the data
particularly satisfactory.

Mouri \& Taniguchi (1995) consider the 1.5--2.5\mic\ spectrum of the
starburst galaxy NGC 6240, where a number of H$_2$ lines are evident.
Based on the relative strengths of the 1--0 S(7) and S(9) lines,
compared to the 1--0 S(1) line, they argue that formation pumping, via
associative detachment of H and H$^-$ to form H$_2$, provides an
important contribution to the line intensity.  However the data on
which this is based have low spectral resolution, and suffers from
blending.  Moreover, the data are fit with a model which contains
several components to the 1--0 S(1) line intensity; formation (10\%),
fluorescence (20\%) and thermal excitation (70\%).  In addition, the
v=6--4 Q(1) line at 1.64\mic, observed by Elston \& Maloney (1990) in
this source, and arising from the same level as the 6--4 O(3) line
which we report on in this paper, is weak compared to the 1--0 S(1)
line, less than 10\% its intensity. It has low signal to noise in the
spectrum. Hence, claims for a formation signature in the data need to
be regarded with caution.

Measurements of over 30 high-excitation H$_2$ lines in the reflection
nebula NGC~2023 by Burton et al.\ (1992) lead these authors to
speculate on whether there was a formation component to the emission
from several lines from the v=4 level.  This was based on an excess in
the level column density distribution for these lines compared to
expectations for pure fluorescent emission (and see also McCartney et
al.\ 1999 for an extension to v=6 in this source, for which the excess
may still be apparent).  However, the signal to noise is not
sufficient to be certain that the excess is real, and moreover, might
be mistaken for ortho-to-para ratio variations between lines in
different vibrational levels.

This paper presents images of emission lines from the v=1, 2 and 6
levels in M\,17, as well as in the hydrogen Br $\gamma$ line.  This
includes the first map ever obtained of an H$_2$ emission line from
the v=6 level. While the bulk of the H$_2$ emission is clearly
fluorescent in origin, we argue, based on the different morphology for
the v=6--4 O(3) line to the lower excitation lines, that formation
pumping of H$_2$ provides a significant component to its flux.

\section{OBSERVATIONS AND DATA REDUCTION}
\label{sec:obs}
Two fields in the source Messier 17 were observed using the IRIS
1--2.5\mic\ camera, in conjunction with the University of New South
Wales Infrared Fabry-P\'{e}rot etalon (UNSWIRF, Ryder et al.\ 1998), on
the 3.9-m Anglo-Australian Telescope (AAT).  The data were obtained on
1997 July 21--23. The fields were chosen to centre on the H$_2$
emission in the northern (N) and south-western (SW) ionization bars of
M~17, at $18^h 20^m 42.8^s$, $-16^{\circ} 8' 23''$ and $18^h 20^m
25.3^s$, $-16^{\circ} 11' 27''$ (J2000), respectively.  These are
shown in Fig.~\ref{fig:overlay}, overlaid on a 3-colour near--IR image
of M~17 (1--2.5\mic) obtained by Lada et al.\ (1991). This image
clearly shows the two ionization bars, with blue indicating the
ionized gas (through Paschen lines of hydrogen in the 1.25\mic\ band)
and red mostly indicating the photodissociation region (PDR, through
fluorescently-excited H$_2$ lines in the 2.2\mic\ band).

The UNSWIRF etalon has a FWHM spectral resolution of $\rm \sim 75 \ km
\ s^{-1}$, a pixel size of 0.77\,arcsec and a 100\,arcsec circular
field of view. It is scanned through a spectral line of interest in
order to obtain an emission line image with minimal contamination from
any continuum radiation present in a source.

The H$_2$ 1--0 S(1) (2.1218\mic), 2--1 S(1) (2.2233\mic), 1--0 S(7)
(1.7480\mic) and 6--4 O(3) (1.7326\mic) lines, and the hydrogen
Br$\gamma$ (n=7--4, 2.1661\mic) and Br10 (n=10--4, 1.7367\mic) lines
were imaged in the N field, and just the 2\mic--band lines in the SW
field.  Each of the three 2\mic\ lines was imaged, using appropriate
1\% width blocking filters, by stepping the etalon through the
relevant plate spacings for the line of interest.  All three 1.7\mic\
lines were imaged with the same 1.5\% width blocking filter. For the
weaker 6--4 O(3) line, which had never been imaged before, the
spectral scan was defined to also include the nearby Br10 line. The
hydrogen line provided a wavelength reference point to ensure the
correct plate spacing was used for the 6--4 O(3) line.

The observing sequence with UNSWIRF included 3 on-line settings for
the 2\mic\ lines, and 5 on-line settings for the 1.7\mic\ lines,
equally spaced in 39 \kms\ steps about the respective line centres.
An `off-line' setting was chosen to provide continuum subtraction.
Sky frames were also taken for each etalon spacing.  An integration
time of two minutes was used for each setting at 2\mic, and five
minutes at 1.7\mic. The corresponding sky frames were taken
immediately after each source frame for each separate Fabry-P\'{e}rot
etalon setting (and located 5\,arcmin N for the N--field and 5\,arcmin
W for the SW--field). One such sequence of frames was taken for the
1--0 S(1) and Br$\gamma$ lines, four for the 2--1 S(1) line (two in
the SW field) and two for the 1.7\mic\ lines.  Each repeated sequence
was centred on a slightly different source position, to minimise the
effect of any bad pixels.

The stars BS7330 (K=4.97 mag) and BS6748 (K=4.57 mag) were used as
flux standards, and were imaged at each of the etalon spacings. A
diffused dome lamp provided a flat field for each etalon spacing, and
an arc lamp was scanned through a free spectral range in order to
wavelength calibrate the Fabry-P\'{e}rot etalon response for each
pixel of the array.

Data reduction was through a custom software package using {\small
IRAF}\footnote{Image Reduction and Analysis Facility
(www.iraf.noao.edu)}.  Frames are linearised, flat-fielded using a
dome flat at the appropriate plate setting, sky-subtracted, shifted to
align stars in each frame, smoothed and the off-line frame subtracted
from each on-line frame (having been appropriately scaled to minimise
residuals from the subtraction process).  Stacking of the frames
yields a data cube, which is then fitted pixel-by-pixel with the
instrumental profile (a Lorentzian) to yield a line image.  Results
from different sequences on the same line were then coadded.  The
absolute accuracy in flux calibration is typically around 30\% for
each image, and line ratios determined from them can thus have a
(constant) scaling error of up to a factor of two. However, the error
in relative line ratios determined between any two pixels is much
smaller than this, depending only on the S/N each line has been
measured with, and not on the accuracy of the absolute calibration.

The line centre for each pixel is determined from the plate spacing
found for peak of the fitted profile, with an accuracy of $\rm \sim 10
\ km \ s^{-1}$ (depending on the intensity).  Within these errors, the
emission velocity for each line across the fields was found to be the
same, as expected for a photodissociation region, where gas motions
are very much less than the spectral resolution of the data.

\begin{figure*} 			 
\begin{center} 
\begin{tabular}{c}
\end{tabular} 
\caption{Three-colour infrared image of M~17 (Lada et al.\ 1991), 
overlaid with circles indicating the two 100\,arcsec fields observed
in the N and SW ionization bars of the source.  The infrared image is
coded in blue for J (1.25\mic), green for H (1.65\mic) and red for K
(2.2\mic) bands. A scale bar indicates 5 arcminutes. Extended emission
in blue mainly arises from Paschen lines of hydrogen, whereas that in
red is mostly from H$_2$ lines.}

\label{fig:overlay}
\end{center}
\end{figure*}

\begin{figure*} 			 
\begin{center} 
\begin{tabular}{c}
\end{tabular} 
\caption{Images from the field of the Northern Bar of M~17 of the four 
molecular hydrogen lines. Going clockwise from top left, they are the
1--0 S(1) (2.1218\mic), 2--1 S(1) (2.2233\mic), 6--4 O(3) (1.7326\mic)
and 1--0 S(7) (1.7480\mic) transitions.  Images are overlaid with
contours of the same lines. Contour levels are as follows: for the
1--0 and 2--1 S(1) lines, starting from, and in steps of $\rm 3 \times
10^{-16} erg
\ s^{-1} cm^{-2} arcsec^{-2}$; for the 6--4 O(3) line, starting from,
and in steps of $\rm 0.5 \times 10^{-16} erg \ s^{-1} cm^{-2}
arcsec^{-2}$; for the 1--0 S(7) line, starting from, and in steps of
$\rm 1 \times 10^{-16} erg \ s^{-1} cm^{-2} arcsec^{-2}$.}

\label{fig:nlines}
\end{center}
\end{figure*}

\begin{figure*} 			
\begin{center} 
\begin{tabular}{c}
\end{tabular} 
\caption{Images from the field of the Northern Bar of M~17 
comparing the distribution of various emission lines.  In all images
the H$_2$ 1--0 S(1) line is shown in greyscale overlaid with the
following contours: 2--1 S(1) (top-left), 6--4 O(3) (top-right), 1--0
S(7) (bottom-right) and H Br$\gamma$ (bottom-left). Contour levels for
the images are as follows: 2--1 S(1), starting from, and in steps of
$\rm 3 \times 10^{-16} erg \ s^{-1} cm^{-2} arcsec^{-2}$; 6--4 O(3)
starting from, and in steps of $\rm 0.3 \times 10^{-16} erg \ s^{-1}
cm^{-2} arcsec^{-2}$; 1--0 S(7), starting from, and in steps of $\rm 1
\times 10^{-16} erg \ s^{-1} cm^{-2} arcsec^{-2}$; Br$\gamma$, starting
from 100, and in steps of 30 $\rm \times 10^{-16} erg \ s^{-1} cm^{-2}
arcsec^{-2}$.  Numbered boxes refer to the apertures used in
Table~\ref{table:fluxes}. }
\label{fig:nratios}
\end{center}
\end{figure*}

\begin{figure*} 			 
\begin{center} 
\begin{tabular}{c}
\end{tabular} 
\caption{Images from the field of the South Western Bar of M~17 
of the 1--0 S(1) (2.1218\mic) (top left) and 2--1 S(1) (2.2233\mic)
(top right) molecular hydrogen lines and the hydrogen Br$\gamma$
(2.1661\mic) line (bottom left). Contours of the line intensity are
overlaid greyscale representations in each of three images.  In the
bottom right image, contours of Br$\gamma$ line emission are overlaid
on a greyscale image of the H$_2$ 1--0 S(1) line.  Contour levels for
the images are as follows: for the 1--0 and 2--1 S(1) lines, starting
from, and in steps of $\rm 2 \times 10^{-16} erg \ s^{-1} cm^{-2}
arcsec^{-2}$; for the Br$\gamma$ line, starting from, and in steps of
$\rm 3 \times 10^{-14} erg \ s^{-1} cm^{-2} arcsec^{-2}$; for the
Br$\gamma$ line overlaid on the H$_2$ image, starting from 1500, and
in steps of 300, $\rm \times 10^{-16} erg \ s^{-1} cm^{-2}
arcsec^{-2}$. Numbered boxes refer to the apertures used in
Table~\ref{table:fluxes}.}
\label{fig:swlines}
\end{center}
\end{figure*}

\section{RESULTS}
\label{sec:results}
\subsection{Northern Bar}
\subsubsection{H$_2$ 1--0 S(1), 1--0 S(7) and 2--1 S(1) Lines}
Fig.~\ref{fig:nlines} shows images of the H$_2$ 1--0 S(1), 2--1 S(1),
6--4 O(3) and 1--0 S(7) lines from the N field.  Line fluxes and line
ratios for selected regions are shown in Table~\ref{table:fluxes}.
Fig.~\ref{fig:nratios} compares, for the N field, the distribution of
the H$_2$ 1--0 S(1) line (in greyscale) to that of the 2--1 S(1), 6--4
O(3), 1--0 S(7) and H Br$\gamma$ lines (in contours), respectively.
The ionization front between the \hii\ and H$_2$ emitting gas can be
seen, though since foreground ionized gas envelops the whole region,
it renders its location hard to discern.

Giving regard to the 30\% error in absolute calibration, discussed in
\S\ref{sec:obs}, the intensity of the 1--0 S(1) line at the peak, $\rm
5 \times 10^{-13} erg \ s^{-1} \ cm^{-2}$ in an $18 \times 15$\,arcsec
aperture, compares reasonably with the flux measured in the N--Bar by
both Tanaka et al.\ (1989) ($\rm 3 \times 10^{-13} erg \ s^{-1} \
cm^{-2}$ in a 20\,arcsec beam) and by Chrysostomou et al.\ (1993)
($\rm 3.5 \times 10^{-13} erg \ s^{-1} \ cm^{-2}$ in long slit with
180\,arcsec$^2$ effective area).  The total 1--0 S(1) line emission
from the N--Bar field is $\rm 5.3 \times 10^{-12} erg \ s^{-1} \
cm^{-2}$, and the brightest emission peaks at $\rm 3.2 \times 10^{-15}
erg \ s^{-1} \ cm^{-2} arcsec^{-2}$ at $18^h 20^m 42.3^s$,
$-16^{\circ} 08' 51''$ (J2000).

The overall morphology of the line emission is one of clumpy
filaments, with two roughly parallel features running SE--NW across
the field.  They are embedded in diffuse H$_2$ emission, extending to
the NE of the field. The morphology is similar in the 1--0 S(1), 2--1
S(1) and 1--0 S(7) line images.  This indicates that the 1--0/2--1
S(1) ratio is reasonably constant across the field.

The line ratio variations are quantified in Table~\ref{table:fluxes}.
The 1--0~S(1)/2--1~S(1) ratio varies from 1.3 to 2.1, with a mean of
1.6.  The 1--0~S(1)/1--0~S(7) ratio varies from 3.9 to 4.9, with a
mean of 4.0.

Chrysostomou et al.\ (1993) also found little variation in the
1--0~S(1)/2--1~S(1) ratio, with a value of $3 \pm 0.5$ over most of
the region they observed.  Tanaka et al.\ (1989) determined 1.9 for
this ratio.  Our own determination is 1.6.  However, given the
uncertainty in absolute calibration for measurements made through
Fabry-P\'{e}rot etalons, these determinations are consistent with one
another.  The value of 3 is more likely to be correct, though, since
measurements through a long slit have better relative accuracy for
line ratio determinations.  This suggests our 1--0~S(1)/2--1~S(1) line
ratios might need to be scaled by $\sim 1.8$ for the N--Bar.

The similarity of the 1--0 S(1) and 1--0 S(7) line images indicates
that any differential extinction between their emitting wavelengths,
1.75 and 2.12\mic, must be uniform (or minimal).  This is consistent
with the estimate of the extinction of 0.05 mag at 2.2\mic\ found by
Chrysostomou et al.\ (1993), based on hydrogen recombination lines at
1.09 and 2.16\mic. For this value of extinction, the differential
extinction between 1.7 and 2.1\mic\ is indeed negligible.

\subsubsection{H$_2$ v=6--4 O(3) Line}
For the v=6--4 O(3) line, while the emission clearly arises from the
same region as the lower excitation lines, its distribution within
that region is quite different.  The line is considerably brighter in
the NE of the two emitting filaments, contrary to what is seen in the
v=1 and v=2 line images.  The number of distinct clumps is also
greater in the 6--4 O(3) image. 

Since the v=6--4 O(3) line is emitted at a similar frequency to the
1--0 S(7) line there will not be any differential extinction between
these two lines. Thus, the difference in morphology between the 6--4
O(3) line and the 1--0 S(1) line reflects differences in their
excitation. It is not the result of differential extinction between
the emitting wavelengths of these two lines.

The 1--0~S(1)/6--4~O(3) ratio varies from 5 to 23 in the N--Bar, with
a mean value of 11 (see Table~\ref{table:fluxes}).  This is a factor
of two variation from the mean ratio, compared to less than 30\%
variation measured for the ratio of the 1--0~S(7) and 2--1~S(1) lines
with the 1--0 S(1) line.

\subsection{South Western Bar}
In Fig.~\ref{fig:swlines} are shown the H$_2$ 1--0 and 2--1 S(1) lines
and the H Br$\gamma$ line for the SW field, together with an overlay
of the Br$\gamma$ line on the 1--0 S(1) line.  Line fluxes are again
listed in Table~\ref{table:fluxes}. The H Br10 line is not shown, but
is similar to the Br$\gamma$ line in appearance.  Again the H$_2$
emission is distributed along clumpy filaments, and (giving regard to
the lower S/N in the v=2 image) the v=1 and 2 lines show similar
morphologies.  A clear ionization front is evident between the
Br$\gamma$ and 1--0 S(1) images, with the brightest emission filaments
running parallel and separated by $\sim 5$\,arcsec.

The total 1--0 S(1) line flux from the SW field is $\rm 2.6 \times
10^{-12} erg \ s^{-1} \ cm^{-2}$. The flux peaks at $\rm 1.9 \times
10^{-15} erg \ s^{-1} \ cm^{-2} arcsec^{-2}$ at $18^h 20^m 23.1^s$,
$-16^{\circ} 11' 16''$ (J2000).  Both these fluxes are about half the
corresponding values in the N--Bar.  This is probably mostly due to
the extra extinction in the optically obscured SW--Bar.  

The 1--0~S(1)/2--1~S(1) line ratio varies between 3 and 8, with a mean
value of 4.4.  This is both higher than in the N--Bar, and shows
significantly more variation with position.  This line ratio cannot be
considered as constant over the field, unlike in the N--Bar.

\section{DISCUSSION}
\subsection{PDR Emission and Gas Density}
As has been discussed in several previous papers, the molecular
hydrogen line emission in M\,17 is dominated by UV fluorescence (e.g.\
Tanaka et al.\ 1989, Chrysostomou et al.\ 1992, Chrysostomou et al.\
1993).  This is evident through the morphology, its proximity to the
ionization front of an \hii\ region, the similarity to UV--excited
3.3\mic\ PAH emission, and through the measurements made of H$_2$
vibrational line ratios.  Our data confirm this result.  In
particular, the strength of the v=2--1 S(1) line compared to the
v=1--0 S(1) line is indicative of a non-thermal excitation method such
as UV fluorescence (e.g.\ Black \& Dalgarno 1976, Black \& van
Dishoeck 1987).

In the SW--Bar the line ratio rises above the pure fluorescent value,
indicating that the density of some of the gas there is greater than
critical ($\sim 10^4$--$10^5$\cmtwo, depending on uncertain values for
the H$_2$ collisional excitation rates). In such cases collisions can
re-populate the v=1 level so that the ratio of the v=1--0 to the
v=2--1 S(1) lines can appear thermal (i.e.\ $\sim 10$, see Sternberg
\& Dalgarno 1989, Burton, Hollenbach \& Tielens 1990).  However when
such ``collisional fluorescence'' is seen, density variations within
the emitting region invariably give rise to significant variations in
the value of this line ratio within the source (see Ryder et al.\
1998, Allen et al.\ 1999).  Density variations must occur in the
SW--Bar of M\,17 too.

However the bulk of the gas in the SW--Bar must have density no more
than critical.  This can be determined from the separation between the
ionization front and the excited H$_2$ there, $\sim 5$\,arcsec.
Assuming this corresponds to an extinction of $A_v \sim 1$ mag (i.e.\
$\rm N \sim 10^{21}$ H$_2$ molecules cm $^{-2}$), it implies that the
average H$_2$ number density is $\rm n \sim 1 \times 10^{4} \ cm^{-3}$
in the SW--Bar.  Given the high density component that is also there,
the medium must therefore be a clumpy one.

In the N--Bar we will take the line ratio to be 3, as determined by
Chrysostomou et at.\ (1993).  This is only a little higher than the
pure fluorescent value, and suggests that the density here is about
equal to the critical density.  The constancy of the ratio across the
N--Bar also suggests that there is also little variation in the
density.  This is in contrast to the clumpy SW--Bar.

\subsection{Molecular Hydrogen Formation Pumping}
The most interesting result from this work is the v=6--4 O(3) emission
line image of the N--bar, shown in Fig.~\ref{fig:nlines}.  This is the
first time an image in such a high-excitation line of H$_2$, 31,000\,K
above ground state, has been obtained.  While the line arises from the
same regions as the v=1 and v=2 lines of H$_2$, its distribution
within them is clearly different.

Since all the lines arise from ortho (odd--J) states, the difference
between them is also unlikely to result from ortho-to-para ratio
variations, unless these both vary with position in the source, as
well as between vibrational levels.

It is also hard to see how the 6--4 O(3) line could have been
mis-identified.  It cannot be a line emitted from the \hii\ region due
to the completely different morphology from the hydrogen lines. Since
the line is emitted from the same regions as the lower excitation
H$_2$ lines it also suggests that it is a PDR line, and not from the
\hii\ region.  Moreover, the nearby H Br10 line provided a wavelength
reference point for scanning the Fabry-P\'{e}rot etalon across the
correct plate spacings for the 6--4 O(3) line.

The v=6--4 O(3) line emission almost certainly arises from a region
where the H$_2$ emission is dominated by UV--fluorescence.  It cannot
be shocked or X--ray excited since the high energy of the (v,J) =
(6,1) upper level (31,000\,K) would not be significantly populated by
thermal means.  The 6--4 O(3) line would be $\sim 10^{-4}$ of the
strength of the 1--0 S(1) line for thermal excitation at 2000\,K,
considerably weaker than the nearby 1--0 S(7) line, which would be
about 10\% of the 1--0 S(1) line strength. In fluorescent models,
however, the v=6--4 O(3) line intensity is significant; for instance
in Black \& van Dishoeck's Model 14 it is 31\% of the value of the
1--0 S(1) line.  In the same model the 2--1 S(1) line is 56\% the
strength, and the 1--0 S(7) line 18\%, of the 1--0 S(1) line intensity
(neglecting any differential extinction between the emitting
wavelengths).  This is broadly consistent with the data, for which the
mean fluxes of the 2--1 S(1), 1--0 S(7) and 6--4 O(3) lines are
$\sim$60\%, 25\% and 10\% of the 1--0 S(1) line, respectively. However
the specific predictions for the strength of the lines depend upon an
additional excitation mechanism to the fluorescent cascade, formation
pumping.

For every H$_2$ molecule that is fluorescently excited, 15\% lead to
photodissociation (e.g.\ Draine \& Bertoldi 1996).  In steady state
PDR models, molecule destruction is balanced by molecule formation,
which is believed to occur on grain surfaces (e.g.\ Hollenbach \&
Salpeter 1971).  The newly formed H$_2$ molecules are released in an
excited vibrational-rotational state.  However, there is little hard
evidence to suggest in what state, or distribution of states, this
might be.  Black \& van Dishoeck (1987) consider three possible
formation models.  Their `standard' Model 14 assumes that 1.5\,eV of
the binding energy is distributed in a Boltzmann distribution through
the energy levels (after Black \& Dalgarno 1976).  A second model
assumes that the new molecules all appear in v=14, J=0 or 1 (after
Hunter \& Watson 1978).  The third formation model is based on a
treatment of H$_2$ catalysis by Duley \& Williams (1986) which
predicts the molecules are ejected into the v=6 level.  Naturally,
there are significant differences in the predictions of the
intensities for lines from v=6 between the first two and the third
formation model.  Depending on the formation temperature of the
molecule, v=6 lines are predicted to be up to a factor $\sim 3$ times
brighter in the third than in the first formation model. This suggests
that the 1--0~S(1)/6--4~O(3) ratio may fall from $\sim 10$ to $\sim 3$
in Black \& van Dishoeck's (1987) Model 14, if it were to be modified
so that H$_2$ formation occured in v=6, rather than into a Boltzmann
distribution.

We believe that formation pumping into v=6, or to a nearby level,
provides the best explanation for the difference in the images between
the 6--4 O(3) line and the lower excitation lines.  However, it cannot
simply be due to the additional component that formation pumping adds
to the intensity of a line, for that would not result in a different
morphology for the line.  Either the rate at which formation is
occurring must vary with position, or the formation spectrum itself is
varying with position (for instance, due to a varying formation
temperature), across the N--Bar of M\,17\@.

In the Duley \& Williams model, about 1.5\,eV of the 4.5\,eV bond
energy released is assumed to remain with the H$_2$ molecule, as it is
ejected from the surface of the dust grain where it formed.  This puts
it into an excited rotational-vibrational state, namely $\rm v \sim
6$.  The actual formation process depends intimately on the nature of
the surface of the dust grains.  In their 1986 paper Duley \& Williams
considered the surfaces to be highly defected silicates, to which the
molecule is moderately strongly bound.  A later paper (Duley \&
Williams 1993) further considers H$_2$ formation on amorphous H$_2$O
ice (where it is weakly bound) and on aromatic carbon molecules (e.g.\
PAHs or HACs, where it is strongly bound).  If the former occurs, they
suggest that the H$_2$ will be released in highly-excited states
(i.e.\ the binding energy virtually all goes into internal energy in
the H$_2$ molecule, so it ends up, perhaps, in v=13). In the latter
case, most of the binding energy would be distributed through the many
degrees of freedom of the aromatic molecule, leaving the H$_2$ in a
low-excitation state.  Clearly, the observations in M\,17 support the
silicate grains model for that source over the other two grain models,
since the excess appears to be in v=6.

The PDR models discussed above were steady state models, where the
dissociation front is stationary. In them, the photodissociation rate
of hydrogen molecules by far--UV radiation is balanced by their
formation rate on grains.  This need not be the case. The timescale
for the H$_2$/H dissociation front to reach equilibrium, $\rm t_{eq}
\sim 5 \times 10^8/n$ years, can be long compared to variations in the
radiation field (for instance, soon after a star switches on)
(Hollenbach \& Natta 1995).  In M\,17, the data constrain $\rm t_{eq}$
to be less than $\sim 10^4$ years. The exposure of previously shielded
H$_2$ molecules to the far--UV radiation field, as the dissociation
front moves further into a molecular cloud, increases the column of
fluoresced gas.  Thus, the H$_2$ line intensities rise from their
steady state values.  The 1--0 S(1) line intensity can be elevated by
an order of magnitude (Hollenbach \& Natta 1995). Moreover, since this
emission is dominated by pure fluorescence (collisions can only play a
minor role in redistributing the level populations), the 1--0/2--1
S(1) ratio will be $\sim 2$. Any thermal contribution to the 1--0 S(1)
line has been minimised.

During this time the rate of molecule destruction must exceed that of
formation.  Thus the relative proportion of the formation component to
the fluorescent H$_2$ line intensities will be reduced.  We now
consider whether this explanation can be applied towards explaining
the 6--4 O(3) line emission from the N--Bar of M\,17. In it, the line
is brighter in the NE of the two emission filaments, the one furthest
away (at least in projected distance) from the ionizing stars in the
\hii\ region.  Our hypothesis is that a formation component
contributes to the 6--4 O(3) line intensity.  This would imply that,
in the weaker of the two filaments, a smaller proportion of the
emission has been produced by formation pumping, compared to that
produced by pure fluorescence, than in the brighter of the filaments.
Thus, this suggests that the SW of the two filaments contains a
non-steady state photodissociation region, where the dissociation
front is moving rapidly into the molecular cloud.

We can apply this interpretation to provide a rough estimate of the
fraction of the 6--4 O(3) line intensity that derives directly from
H$_2$ formation in a steady state PDR\@.  We assume that when the
1--0~S(1)/6--4~O(3) ratio is at it largest (i.e.\ 23), then the UV
cascade dominates its excitation.  We also assume that when it is at
its smallest value (i.e.\ 5) this represents the steady state PDR\@.
This then yields a formation component that is $\sim 80$\% of the
total 6--4 O(3) line intensity in the steady state, the remaining
$\sim 20$\% coming from the fluorescent cascade.  This is obviously a
crude estimate.  Without further data on other high excitation H$_2$
lines we cannot say more about the formation spectrum of H$_2$.


\section{SUMMARY}

The molecular hydrogen line emission from the N-- and SW--Bars of
M\,17 clearly is fluorescently excited, from inside a photdissociation
region. The N--Bar has approximately constant density, whereas the
SW--Bar contains a clumpy molecular medium.  However, the different
distribution of the emission from the v=6--4 O(3) line, compared to
that of the lower excitation lines from the v=1 and 2 levels,
indicates that another mechanism is significant for its excitation.
We argue that this is formation pumping, newly formed molecules being
released from grain surfaces in an excited state, at or near the v=6
level.  This therefore resembles the suggestion put forward by Duley
\& Williams (1986) for H$_2$ formation on the surface of highly
defected silicate grains, where the molecules are released with $\sim
1.5$\,eV of the 4.5\,eV molecule binding energy.  However,
time-dependent PDR models are also needed in order to quantitatively
model such emission.  This is because the proportion of the H$_2$ line
emission that has been produced by formation pumping, in comparison to
fluorescence, probably varies with position in the source.

If this hypothesis is correct it may be tested by imaging in other
highly excited emission lines.  For instance, the v=6--4 Q(1) line at
1.6011\mic, the 5--3 O(3) line at 1.6131\mic, the 7--5 S(1) line at
1.6201\mic\ and the 7--5 Q(1) line at 1.7283\mic, are all emitted from
nearby levels and in the same atmospheric window.  The v=6 and v=5
lines would likely look similar, but, depending on whether any H$_2$
is pumped into v=7 on formation, the distribution of the latter two
lines may look closer to that of the lower excitation v=1 lines.
Imaging Fabry-P\'{e}rot etalons are suitable instruments for these
observations, because of their high spectral resolution, needed to
separate out other lines in the window, and because they enable the
spatial morphology to be examined.  Long-slit spectroscopy, again at
high spectral resolution, across the emission bars, would permit
further examination of the merits of the formation hypothesis.  It
would help determine which lines have a formation component to their
emission fluxes.  Such data are needed to provide constraints on what
the H$_2$ formation spectrum actually is.

\section*{ACKNOWLEDGEMENTS}

We are grateful to Michael Ashley, Lori Allen, Jung-Kyu Lee, Stuart
Ryder, John Storey and Andrew Walsh, as well as to the staff of the
Anglo Australian Telescope, for all their help and support during the
conduct of the observations.  We also thank Michael Merrill for
providing the beautiful colour image of M\,17, shown in
Fig.~\ref{fig:overlay}.  We also acknowledge the support of the
Australian Research Council for the funding to undertake this
research.

\begin{table*}

\caption{Molecular hydrogen line fluxes in selected regions of the N--Bar 
and the SW--Bar of M\,17.  The regions are numbered and shown in
Fig.~\ref{fig:nratios} and Fig.~\ref{fig:swlines}, respectively.  The
coordinates (in J2000) of the brightest emission in each region is
listed in columns 2--3. The aperture size, in arcsec, is indicated in
column 4.  In columns 5--8 line fluxes, in $\rm 10^{-14} erg \ s^{-1}
\ cm^{-2}$, are listed for the 1--0 S(1), 2--1 S(1), 1--0 S(7) and
6--4 O(3) lines.  Line ratios for 1--0 S(1)/2--1 S(1), 1--0 S(1)/1--0
S(7) and 1--0 S(1)/6--4 O(3) are listed in columns 9--11.  At the end
of the tabulation for each Bar, the line fluxes and ratios for the
whole field are also listed. }

\begin{tabular}{ccccccccccc}
\hline
Region & RA & Dec & Aperture & 1--0 S(1) & 2--1 S(1) & 1--0 S(7) & 6--4 O(3) & 1--0 S(1)
& 1--0 S(1) & 1--0 S(1) \\ 
& $18^h 20^m$ & $-16^{\circ}$ & arcsec & \multicolumn{4}{c}{$\rm
\times 10^{-14} erg \ s^{-1} \ cm^{-2}$} & / 2--1 S(1) & / 1--0 S(7) &
/ 6--4 O(3) \\
\hline 
\multicolumn{9}{l}{Northern Bar} \\
1 & $43.6^s$ & $08' 30''$ & $18 \times 15$ & $48.4 \pm 0.6$ & $29.5 \pm 0.6$ & $12.1 \pm 0.2$ & $3.4 \pm 0.1$ & $1.6 \pm 0.1$ & $4.0 \pm 0.1$ & $14.1 \pm 0.5$  \\ 
2 & $41.4^s$ & $08' 35''$ & $18 \times 17$ & $30.1 \pm 0.5$ & $15.6 \pm 0.5$ & $7.7  \pm 0.2$ & $1.8 \pm 0.1$ & $1.9 \pm 0.1$ & $3.9 \pm 0.2$ & $17.2 \pm 0.9$  \\ 
3 & $42.3^s$ & $08' 51''$ & $17 \times 13$ & $24.5 \pm 0.4$ & $11.8 \pm 0.4$ & $6.1  \pm 0.1$ & $1.1 \pm 0.1$ & $2.1 \pm 0.1$ & $4.0 \pm 0.2$ & $22.7 \pm 1.2$  \\
4 & $42.8^s$ & $07' 50''$ & $9 \times 12$  & $12.5 \pm 0.4$ & $8.4  \pm 0.4$ & $3.0  \pm 0.1$ & $1.1 \pm 0.1$ & $1.5 \pm 0.1$ & $4.1 \pm 0.3$ & $11 \pm 1$  \\ 
5 & $45.3^s$ & $08' 33''$ & $15 \times 16$ & $25.1 \pm 0.6$ & $19.2 \pm 0.6$ & $6.2  \pm 0.2$ & $3.8 \pm 0.1$ & $1.3 \pm 0.1$ & $4.1 \pm 0.2$ & $6.6 \pm 0.4$  \\ 
6 & $43.6^s$ & $07' 51''$ & $12 \times 15$ & $15.4 \pm 0.5$ & $11.4 \pm 0.5$ & $3.2  \pm 0.2$ & $3.1 \pm 0.1$ & $1.4 \pm 0.1$ & $4.9 \pm 0.5$ & $5.0 \pm 0.4$  \\ 
Total &&                  & $83 \times 85$ & $528 \pm 3$ & $335 \pm 3$   & $132  \pm 1$   & $49  \pm 0.5$ & $1.6 \pm 0.1$ & $4.0 \pm 0.1$ & $10.9 \pm 0.2$  \\ 
\\
\multicolumn{9}{l}{South Western Bar} \\
1 & $24.9^s$ & $11' 37''$ & $15 \times 19$ & $34.8 \pm 0.7$ & $4.4 \pm 0.7$ &&& $7.9 \pm 1.4$ &&  \\ 
2 & $26.3^s$ & $11' 04''$ & $14 \times 13$ & $12.6 \pm 0.5$ & $4.0 \pm 0.5$ &&& $3.1 \pm 0.5$ &&  \\ 
3 & $23.1^s$ & $11' 16''$ & $11 \times 12$ & $17.3 \pm 0.4$ & $6.1 \pm 0.4$ &&& $2.8 \pm 0.3$ &&  \\ 
4 & $23.4^s$ & $12' 00''$ & $15 \times 21$ & $19.9 \pm 0.7$ & $3.9 \pm 0.7$ &&& $5   \pm 1$   &&  \\ 
Total &&                  & $66 \times 84$ & $264 \pm 3$ & $61 \pm 3$   &&& $4.4 \pm 0.3$ &&  \\ 
\hline

\end{tabular}
\label{table:fluxes}
\end{table*}

\bibliographystyle{mnras}

\label{lastpage}

\end{document}